\documentclass[twocolumn,10pt]{article}
\usepackage{times}
\usepackage{amssymb}
\usepackage{amsthm}
\usepackage{url}

\usepackage[dvips]{graphicx}
\title{Some protein interaction data do not exhibit
power law statistics}
\author{Reiko Tanaka$^1$\footnote{Corresponding author:
reiko@bmc.riken.jp,
Bio-Mimetic Control Research Center, RIKEN,
Nagoya 223-8522, Japan}, Tau-Mu Yi$^2$ and John Doyle$^3$}
\date{\small 1 Bio-Mimetic Control Research Center, RIKEN\\
2 Developmental and Cell Biology,
University of California, Irvine\\ 
3 Control and Dynamical Systems,
 California Institute of Technology}

\begin{document}
\maketitle
\begin{abstract}
It has been claimed that protein-protein interaction (PPI)
networks are scale-free based on the observation that the
node degree sequence follows a power law. Here we argue
that these claims are likely to be based on erroneous
statistical analysis. Typically, the supporting data are
presented using frequency-degree plots. We show that such
plots can be misleading, and should correctly be replaced
by rank-degree plots. We provide two PPI network examples
in which the frequency-degree plots appear linear on a
log-log scale, but the rank-degree plots demonstrate that
the node degree sequence is far from a power law. We
conclude that at least these PPI networks are not scale-free.\\
\end{abstract}


\noindent{\bf List of abbreviations}
PPI: Protein-protein interaction, SF: Scale-free, SR: Scale-rich

\section{Introduction}

Experimental data on protein-protein interaction (PPI)
networks have been extensively gathered with the aim of
acquiring a system-level understanding of biological
processes~\cite{Uetz,Vidal}. Various statistical features
of complex graphical structures have received attention,
including the size of the largest connected component, the
node degree distribution, the graph diameter, the
characteristic path length, and the clustering coefficient.
However, the feature that has attracted the most attention
is the distribution of node degree (the number of links
from a node) and whether or not the distribution follows a
power law (linear plot on log-log scale). The degree
distribution of PPI networks was claimed to follow a power
law in~\cite{lethality}, and thus PPI networks are
considered to be "scale-free" (SF)~\cite{NetBio}, a generic
property of network topologies common to various networks
in different domains, from social networks and biological
systems to the Internet.

Although ``scale-free'' has not been clearly defined in the
existing literature~\cite{NetBio}, most treatments assume
that a power law node degree distribution is an important,
and sometimes defining feature. Other characteristics
described in the SF literature include failure tolerance
but attack vulnerability at hubs (nodes possessing high
degree) and various kinds of self-similarity. A recent
attempt at a more theoretically rigorous
treatment~\cite{SFtheory} shows however that no additional
features follow from power law node degree sequence alone,
and require additional restrictions, such as high
likelihood of occurrence by random generation (e.g. by
preferential attachment). Other work~\cite{Natasa,topology}
has highlighted important differences between PPI networks
and "SF networks" constructed by a stochastic growth model.
Moreover, one may question the rigor with which the power
law node degree distribution, the primary feature of SF
networks, has been demonstrated in certain examples.

This letter shows that the node degree sequences of some
published PPI networks are better
described by an exponential function when properly
plotted and analyzed. The problem with previous work is
that data were plotted using frequency-degree plots, as is
common in papers purporting to discover power laws in
complex biological systems, which lead to systematic errors
compared with rank-degree plots. We demonstrate here that
data plotted on a loglog scale frequency-degree plot may
appear to be linear, but when the same data are plotted on
a loglog scale rank-degree plot, they are clearly shown
not to be power law. Thus, the data for some PPI networks
lack even the minimal features of scale-free networks.

\section{Materials and Methods}

Publicly available data for PPI networks represent only an
approximation of the real interaction network because of
the large number of false positive and false negative
interactions. However, because of the assumed
self-similarity features of SF networks, it has been
claimed that if the real PPI network is SF, then any
appropriately sampled subnetwork is also
SF~\cite{Barabasi}. Thus, we might still gain valuable
information by examining whether the publicly available PPI
network data possess a power law node degree distribution
characteristic of SF networks.

A finite sequence of node degrees
$y=(y_1, y_2, \dots, y_n$) of integers,
assumed without loss of generality always to be ordered such that
$y_1 \ge y_2 \ge \ldots \ge y_n$, is said to follow a power law
if
\begin{equation}
    k  \approx  c {y_k}^{-\alpha},
    \label{eq:scaling}
\end{equation}
where $k$ is (by definition) the rank of $y_k$, $c>0$ is a
constant, and $\alpha>0$ is called the scaling index.
Because of the ordering, the rank $k$ is the number of
nodes with the degree equal or larger than $y_k$. Since
$\log k = \log c - \alpha \log y_k$, the rank $k$
versus the node degree $y_k$ plot on a loglog scale appears
as a straight line of slope $-\alpha$. In contrast, $y$ is
said to follow an exponential if
\begin{equation}
    k  \approx  a \exp^{-by_k},
    \label{eq:exponential}
\end{equation}
where $a>0$ and $b>0$ are constants. The $k$ versus $y_k$
plot on a semilog scale approximates a straight line of
slope of $-b$ since $\log k=\log a - by_k$.

Note that the rank-degree relationships (\ref{eq:scaling})
and (\ref{eq:exponential}) are non-stochastic, in the sense
that there need be no assumption of an underlying
probability distribution for the sequence $y$.  Indeed, no
coherent justification has been given for why biological
networks should be viewed as samples from a random
ensemble.  On the contrary, what is known of evolution
would suggest that it yields extremely nonrandom structure
at every level of organization.  Nevertheless, random
graphs have been remarkably popular models for biological
networks, but have led to substantial confusion,
particularly with regard to power laws. Suppose a
non-negative random variable $X$ has cumulative
distribution function (CDF) $F(x)=P[X \leq x]$. In this
stochastic context, a random variable $X$ or its
corresponding distribution function $F$ is said to follow a
power law with index $\alpha>0$ if, as
$x\rightarrow\infty$,
\begin{eqnarray}
 P[X>x] = 1-F(x) \approx cx^{-\alpha},
\label{eq:rv-scaling}
\end{eqnarray}
for some constant $c>0$ and a tail index $\alpha > 0$,
where $f(x) \approx g(x)$ as $x\rightarrow\infty$ if
$f(x)/g(x) \rightarrow 1$ as $x\rightarrow\infty$.
We call (\ref{eq:rv-scaling}) the stochastic
form of power law rank-degree relationship. The
loglog plot of $P[X>x]$ versus $x$ appears as a straight
line of slope $-\alpha$ for large $x$. If the CDF $F(x)$
satisfying (\ref{eq:rv-scaling})
is differentiable, then its derivative, the probability density
function $f(x) = \frac{d}{dx}F(x)$, satisfies
\begin{eqnarray}
f(x) \approx c' x^{-(1+\alpha)}.
 \label{eq:density}
\end{eqnarray}
The loglog plot of $f(x)$ versus $x$ also would be a line
of slope $-(1+\alpha)$. In contrast to the rank-degree
relationships (\ref{eq:scaling}) and
(\ref{eq:exponential}), the definitions in
(\ref{eq:rv-scaling}) and (\ref{eq:density}) are stochastic
and require an underlying probability model. As is standard
in physics, the SF literature almost exclusively assumes
some underlying stochastic models, and power law node
degree distributions are typically investigated in terms of
the frequency-degree relationship based on the probability
density function $f(x)$.

In the case of node degree of graphs the data is inherently
discrete. Even if the data were sampled from some ensemble,
$F(x)$ is not differentiable and the frequency-degree plots
simply do not make sense and can easily lead to mistakes.
Furthermore, differentiation of noisy data, such as PPI
data, amplifies errors, making frequency-based data
uninformative and ambiguous. A typical approach to overcome
these problems is to smooth the data or to group individual
data values into a small number of bins, and then plot the
relative number of data values in each bin. The problem is
that this smoothing or binning process can dramatically
change the nature of frequency-based statistics as will be
shown below (Figs.\ref{fig:human_plots} and
\ref{fig:FYI_plots}). This use of ad hoc statistical
analysis can lead to concluding incorrectly that a power
law relationship is present (or absent). This problem is
easily avoided if one were to make rank-degree plots of raw
data instead of using frequency-degree plots to check the
power law or exponential relationships in
(\ref{eq:scaling}) and (\ref{eq:exponential}).

From among many publicly available studies on PPI networks,
we used the filtered yeast interactome (FYI) data set
\cite{FYI} and the predicted human protein-interaction
(HPI) map \cite{human} to illustrate these points. Much of
the original data suffers from numerous false positives and
false negatives, but more recent investigations have sought
to refine the data. For example, the FYI data set contains
high-confidence interactions for yeast, each observed by at
least two different methods, thereby enriching for genuine
positives. The HPI map was generated using data from seven
experimental and four computationally predicted
protein-interaction maps from {\em Saccharomyces
cerevisiae} \cite{sc1,sc2,sc3,sc4,sc5,sc6}, {\em Drosophila
melanogaster}\cite{dm} and {\em Caenorhabditis elegans}
\cite{ce}. The idea is that a human protein interaction can
be predicted if orthologs in a model organism show an
interaction. Its accuracy has been assessed in
\cite{human}. We consider both FYI and HPI to be refined
data sets, and investigate whether their node degree
sequences follow a power law, a defining feature of
scale-free networks, by rank-degree plots.

\section{Results and Discussion}

\begin{figure}[ht]
  \begin{center}
    \includegraphics[width=0.4\textwidth]{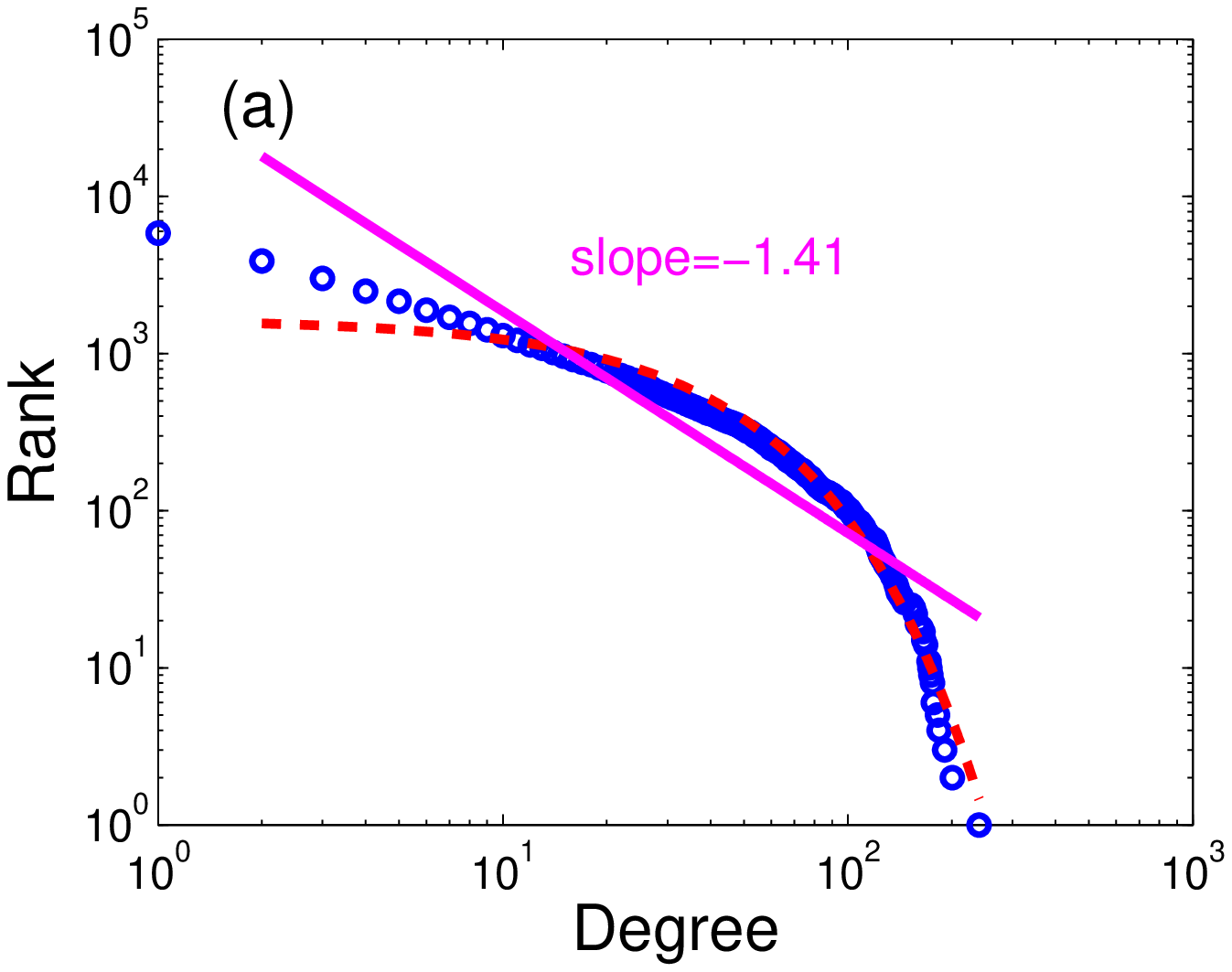}
    \includegraphics[width=0.4\textwidth]{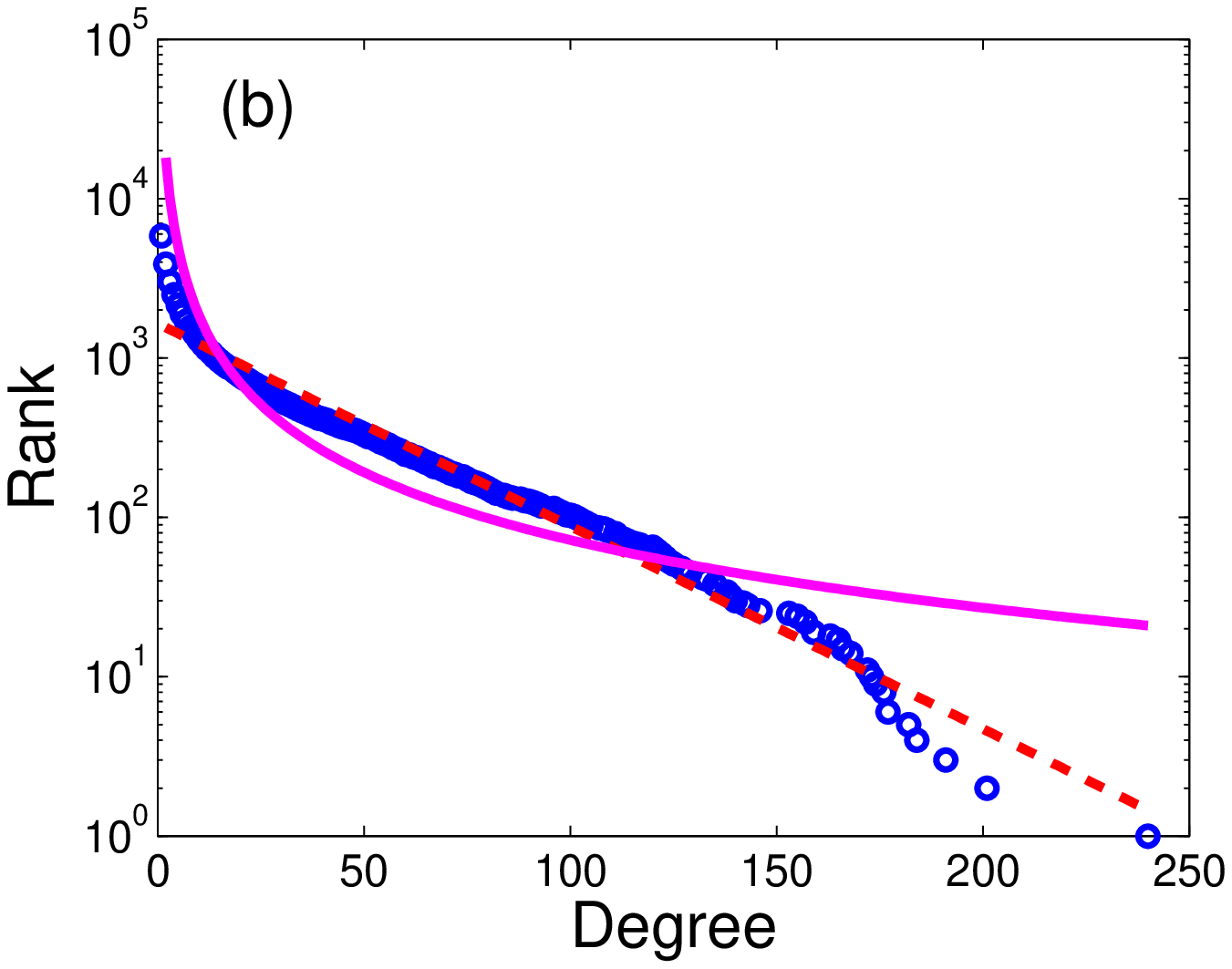}
     \includegraphics[width=0.4\textwidth]{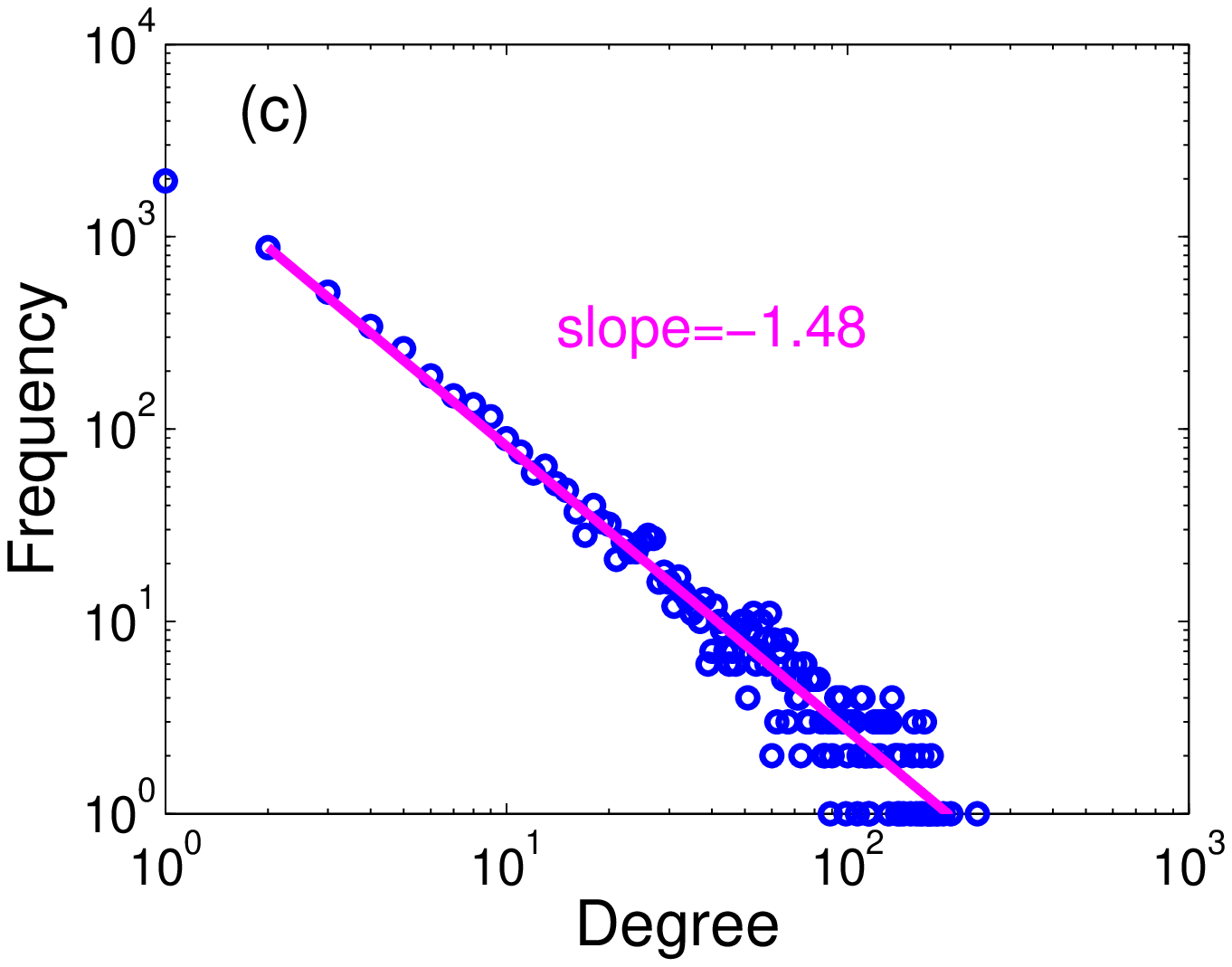}
 \end{center}
\caption{Node degree distribution of human
protein-interaction map \cite{human}: (a) rank-degree plot
in loglog scale, (b) rank-degree plot in semilog scale, and
(c) frequency-degree plot in loglog scale. The rank-degree
plots indicate that the degree distribution is exponential.
The straight lines (a,c) and the dotted curve (a) in loglog
scale are the least-squares fits of the data to the power
law (with the value of the slope) and to the exponential
distributions, respectively. The straight line and the
dotted curve in loglog scale (a) become the curve and the
dotted line in semilog scale (b). Still, the
frequency-degree plot in (c) might appear visually to
follow a power law, and can lead to potential errors of
finding power law node degree distribution.}
  \label{fig:human_plots}
\end{figure}

\begin{figure}[ht]
  \begin{center}
    \includegraphics[width=0.4\textwidth]{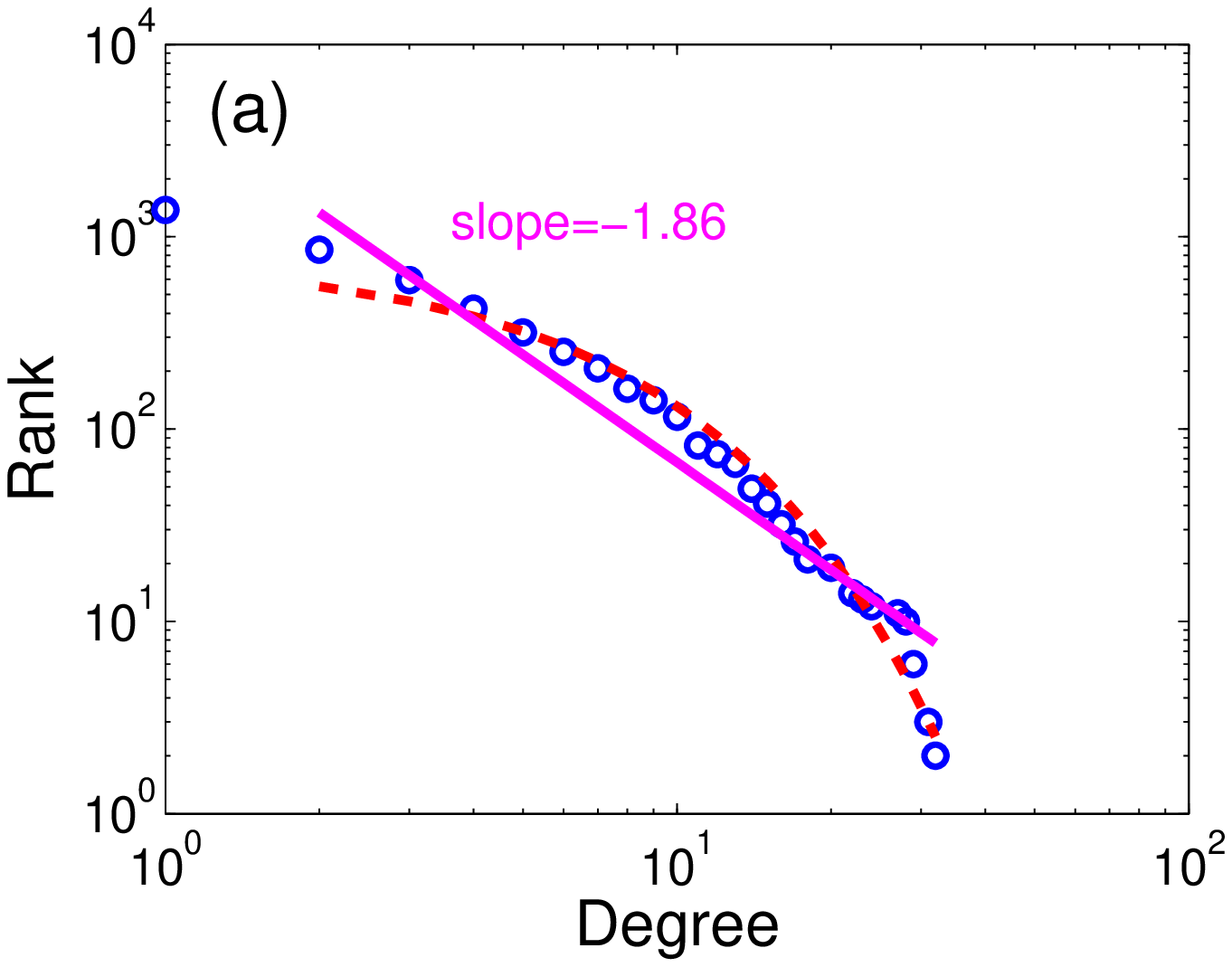}
\includegraphics[width=0.4\textwidth]{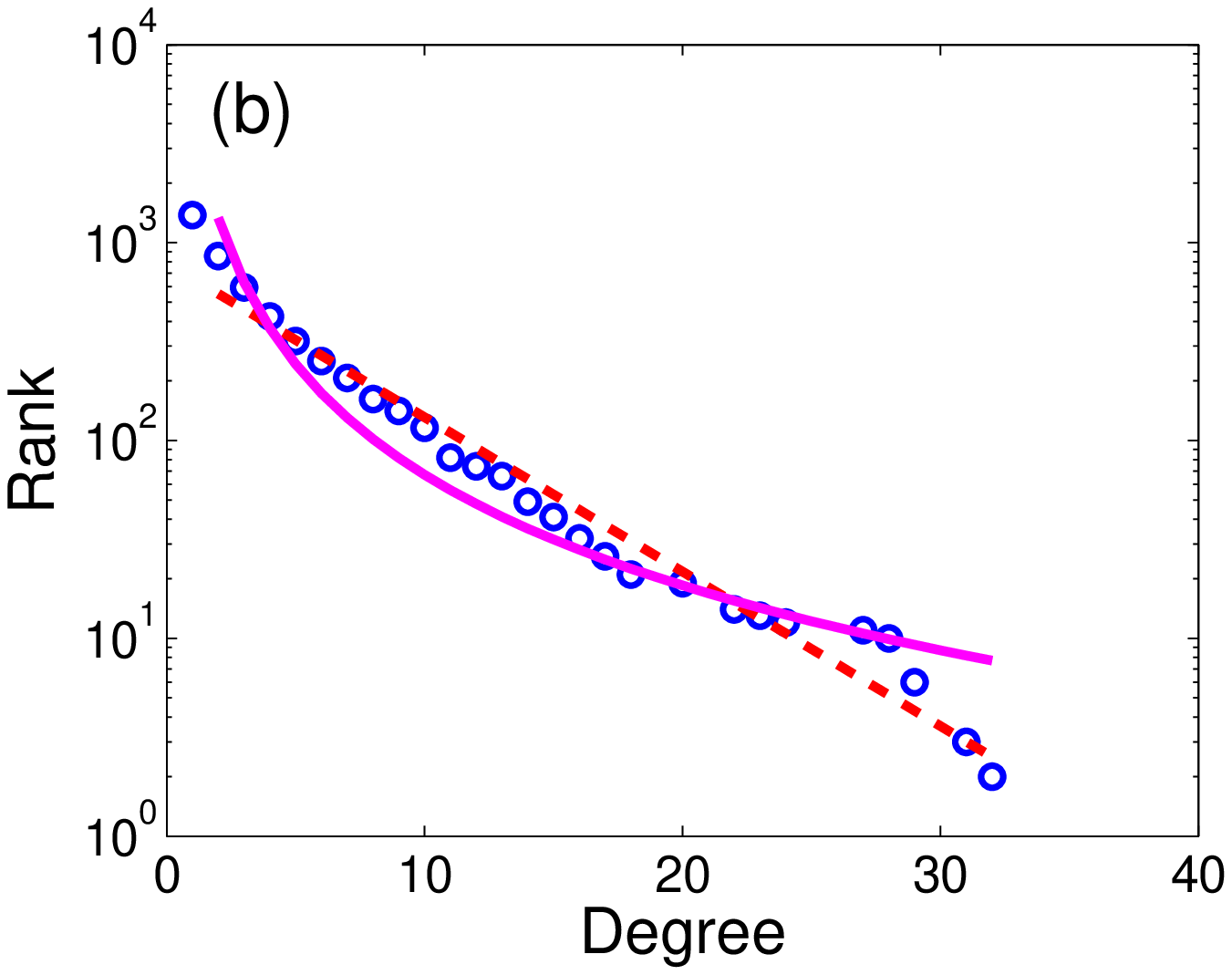}
    \includegraphics[width=0.4\textwidth]{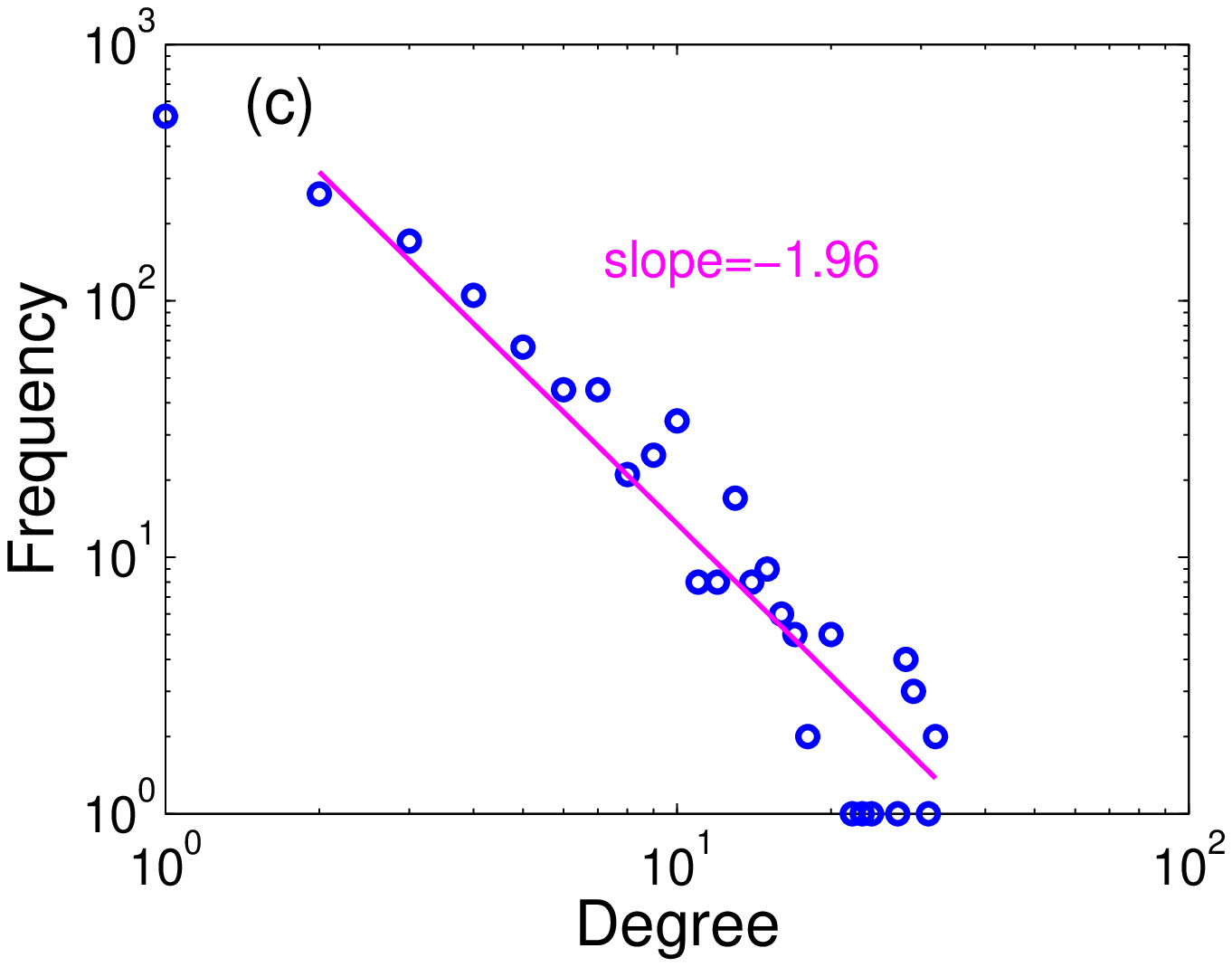}
  \end{center}
  \caption{Node degree distribution of 'filtered yeast interactome'
(FYI) data set \cite{FYI}: (a) rank-degree plot in loglog
scale, (b) rank-degree plot in semilog scale, and (c)
frequency-degree plot in loglog scale. The rank-degree plot
(a,b) shows the non-power law distribution, which is not
evident in the frequency-degree plot (c). The straight
lines (a,c) and the dotted curve (a) in loglog scale are
the least-squares fits of the data to the power law (with
the value of the slope) and to the exponential
distributions, respectively. The straight line and the
dotted curve in loglog scale (a) become the curve and the
dotted line in semilog scale (b).
      }
  \label{fig:FYI_plots}
\end{figure}

The rank-degree plots of the HPI and FYI data are shown in
(a) loglog scale and (b) semilog scale in Figs.
\ref{fig:human_plots} and \ref{fig:FYI_plots},
respectively. The straight lines and the dotted curve in
loglog scale (a) show least-squares fitting of data to a
power law with the value of its slope and to an
exponential, respectively. The same fittings are depicted
as the curve and the dotted straight line in semilog scale
(b). From these figures, we can clearly conclude that the
node degree sequences of HPI and FYI data are much closer
to an exponential (\ref{eq:exponential}), and are clearly
not power laws (\ref{eq:scaling}).  More sophisticated
statistical analysis can be used to confirm these
conclusions.  In addition, the rank-degree plots show raw
data and readers can easily judge at a glance the relative
suitability of various models.

%

However, using frequency-degree plots (c) in
Figs.\ref{fig:human_plots} and \ref{fig:FYI_plots} could
lead to the erroneous conclusion that the node degree
sequence appears to follow a power law, although the
correct rank-degree plot clearly shows that this is not the
case. Furthermore, even if the PPI data were a power law,
the slope for frequency-degree plot $-\beta$ is simply not
related to the slope for the rank-degree plot $-\alpha$ by
$\beta=\alpha+1$, as holds for differentiable
distributions. These results conclusively demonstrate that
these two refined PPI data sets are not power laws, and
thus certainly not scale-free, no matter how this is
defined.

It is in principle possible that the data studied here is
misleading and real PPI networks might have some features
attributed to scale-free networks. At this time we only can
draw conclusions about (noisy) subgraphs of the true
network since the data sets are incomplete and presumably
contain errors. However, the fact that these subgraphs
exhibit an exponential node degree sequences suggests that
the entire network is not SF. Appropriately sampled
subraphs of a SF graph should be SF, and hence possess a
power law node degree sequence. Furthermore, a SF network
possessing significant non-SF subnetworks could not be
considered to be self-similar, a typically assumed though
as yet unproven feature of scale-free networks. Finally,
since essentially all claims that biological networks are
scale-free are based on error-prone frequency-degree
analysis, this analysis must be completely redone to
determine the correct form of the degree sequences.

It has also been shown~\cite{Sigcomm,PRL} that the Internet
and cell metabolism, the two most prominent examples of SF
networks, might have power laws for some degree sequences,
but have none of the other features attributed to
scale-free networks. One important feature of the Internet
and metabolic networks is the complete absence of centrally
located high-degree hubs which are responsible for global
network connectivity and whose removal would fragment the
network, in contrast to what has been claimed in the SF
literature. Metabolic networks have also been shown to be
scale-rich (SR), but not SF, in the sense that they are far
from self-similar \cite{PRL} despite some power laws in
certain node degree sequence. Their power law node degree
sequence is a result of the mixture of exponential
distributions in each functional module. In principle, PPI
networks could have this SR structure as well, and perhaps
power laws could emerge at higher levels of organization.
This will be revealed only when a more complete network is
elucidated. Still, the most important point is not whether
the node degree sequence follows a power law, but whether
the variability of the node degree sequences is high or low
\cite{PRL}, and the biological protocols that necessitate
this high or low variability. These issues will be explored
in future publications.

The authors thank Nicolas Bertin and Marc Vidal for
providing FYI data.

\end{document}